\documentclass[sigconf]{acmart}
\usepackage{enumitem}
\usepackage{subcaption}
\usepackage{numprint}
\usepackage[flushmargin]{footmisc}
\npthousandsep{,}
\usepackage[ruled,linesnumbered]{algorithm2e}
\newcommand{\tabincell}[2]{\begin{tabular}{@{}#1@{}}#2\end{tabular}}  

\makeatletter
\newcommand{\removelatexerror}{\let\@latex@error\@gobble}
\makeatother

\AtBeginDocument{%
  \providecommand\BibTeX{{%
    \normalfont B\kern-0.5em{\scshape i\kern-0.25em b}\kern-0.8em\TeX}}}

\copyrightyear{2021}
\acmYear{2021} 
\setcopyright{iw3c2w3}
\acmConference[WWW '21]{Proceedings of the Web Conference 2021}{April 19--23, 2021}{Ljubljana, Slovenia}
\acmBooktitle{Proceedings of the Web Conference 2021 (WWW '21), April 19--23, 2021,
Ljubljana, Slovenia}

\acmPrice{}
\acmDOI{10.1145/3442381.3449946}
\acmISBN{978-1-4503-8312-7/21/04}

\author{Yongji Wu$^{12*}$, Defu Lian$^{2\dagger}$, Neil Zhenqiang Gong$^{1}$, Lu Yin$^{3}$,\\ Mingyang Yin$^{3}$, Jingren Zhou$^{3}$,  Hongxia Yang$^{3\dagger}$}

\affiliation{
\institution{$^{1}$Duke University, $^{2}$University of Science and Technology of China, $^{3}$Alibaba Group}}
\affiliation{
\institution{$^1$\{yongji.wu769, neil.gong\}@duke.edu, $^2$liandefu@ustc.edu.cn, \\$^3$\{theodore.yl, hengyang.ymy, jingren.zhou, yang.yhx\}@alibaba-inc.com}}

\renewcommand{\shortauthors}{Y. Wu, et al.}
\renewcommand{\authors}{Yongji Wu, Defu Lian, Neil Zhenqiang Gong, Lu Yin, Mingyang Yin, Jingren Zhou, Hongxia Yang}

\begin{document}

\title{Linear-Time Self Attention with Codeword Histogram for Efficient Recommendation}

\begin{abstract}
Self-attention has become increasingly popular in a variety of sequence modeling tasks from natural language processing to recommendation, due to its effectiveness. However, self-attention suffers from quadratic computational and memory complexities, prohibiting its applications on long sequences. Existing approaches that address this issue mainly rely on a sparse attention context, either using a local window, or a permuted bucket obtained by locality-sensitive hashing (LSH) or sorting, while crucial information may be lost. Inspired by the idea of vector quantization that uses cluster centroids to approximate items, we propose LISA (\textbf{LI}near-time \textbf{S}elf \textbf{A}ttention), which enjoys both the effectiveness of vanilla self-attention and the efficiency of sparse attention. LISA scales linearly with the sequence length, while enabling full contextual attention via computing differentiable histograms of codeword distributions. Meanwhile, unlike some efficient attention methods, our method poses no restriction on casual masking or sequence length. We evaluate our method on four real-world datasets for sequential recommendation. The results show that LISA outperforms the state-of-the-art efficient attention methods in both performance and speed; and it is up to 57x faster and 78x more memory efficient than vanilla self-attention.
\end{abstract}



\maketitle
{
\renewcommand{\thefootnote}{\fnsymbol{footnote}}
\footnotetext[1]{This work was conducted while he was a research intern at Alibaba Group.}
\footnotetext[2]{Corresponding authors.}
}

\section{Introduction}
Since the introduction of the self-attention mechanism in Transformers~\cite{vaswani2017attention}, it has seen incredible success in a variety of sequence modeling tasks in a variety of fields, such as machine translation~\cite{chen2018best}, object detection~\cite{wang2018non}, music generation~\cite{huang2018music} and bioinformatics~\cite{madani2020progen}. Recently, self-attention has also demonstrated its formidable power in recommendation~\cite{kang2018self,zhang2018next,sun2019bert4rec}.

However, despite impressive performance attributable to its ability to identify complex dependencies between elements in input sequences, self-attention based models suffers from soaring computational and memory costs when facing sequences of greater length. As a consequence of computing attention scores over the entire sequence for each token, self-attention takes $O(L^2)$ operations to process an input sequence of length $L$. This hinders the scalability of models built on self-attention in many settings.

Recently, a number of solutions have been proposed to address this issue. The majority of these approaches~\cite{kitaev2019reformer,zaheer2020big,roy2020efficient,tay2020sparse,child2019generating,ainslie2020encoding,beltagy2020longformer} leverages sparse attention patterns, limiting the number of keys that each query can attend to. Although these sparse patterns can be established in a variety of content-depended ways like LSH~\cite{kitaev2019reformer}, sorting~\cite{tay2020sparse} and k-means clustering~\cite{roy2020efficient}, crucial information may be lost by clipping the receptive field for each query. While successfully reducing the cost of computing attention weights from $O(L^2D)$ to $O(LBD)$, where $B$ is the fixed bucket size, extra cost incurs in assigning the keys/values into buckets. This cost typically is still quadratic with respect to $L$, and it may cause significant overheads dealing with shorter sequences. We observe that Reformer~\cite{kitaev2019reformer} could be 7.6x slower than the vanilla Transformer on sequences of length 128. Other techniques are also being employed to improve the efficiency of self-attention. For instance, low-rank approximations of the attention weights matrix is used in~\cite{wang2020linformer}. This method, however, only supports a bidirectional attention mode and assumes a fixed length of input sequences.

We observe that self-attention essentially computes a weighted average of the input sequences for each query, and the weights are computed based on the inner product between the query and the keys. For each query, keys with larger inner product will be paid more attention to. We relate this to the Maximum Inner Product Search (MIPS) problem. The MIPS problem is of great importance in many machine learning problems~\cite{koren2009matrix,felzenszwalb2009object,shrivastava2014asymmetric}, and fast approximate MIPS algorithms are well studied by researchers. Among them, vector (product) quantization~\cite{gray1998quantization,guo2016quantization,dai2020norm} has been a popular and successful method. Armed with vector quantization, we no longer have to exhaustively compute the inner product between a given query and all the $N$ points in the database. We can only compute that for the $B$ centroids (i.e., codewords), where $B$ is a budget hyperparameter. We therefore successfully avoid redundant computations since the points belong to the same centroid share the same inner product with the query.

The idea of vector quantization has also been applied to compress the item embedding matrix and improve the memory and search efficiency of recommendation systems~\cite{chen2018learning,lian2020lightrec}. In the state-of-the-art lightweight recommendation model, LightRec~\cite{lian2020lightrec}, a set of $B$ differentially learnable codebooks are used to encode items, each of which is composed of $W$ codewords. An item is represented by a composition of the most similar codeword within each codebook. Hence we only need to store the indices of its corresponding codewords, instead of its embedding vector. Since the codeword index in a codebook can be compactly encoded with $\log W$ bits, the overall memory requirements to store item representations can be reduced from $4ND$ bytes to $\frac{1}{8}NB\log W + 4DBW$ bytes~\cite{lian2020lightrec}.

\begin{figure}
\centering
	\includegraphics[width=.95\linewidth]{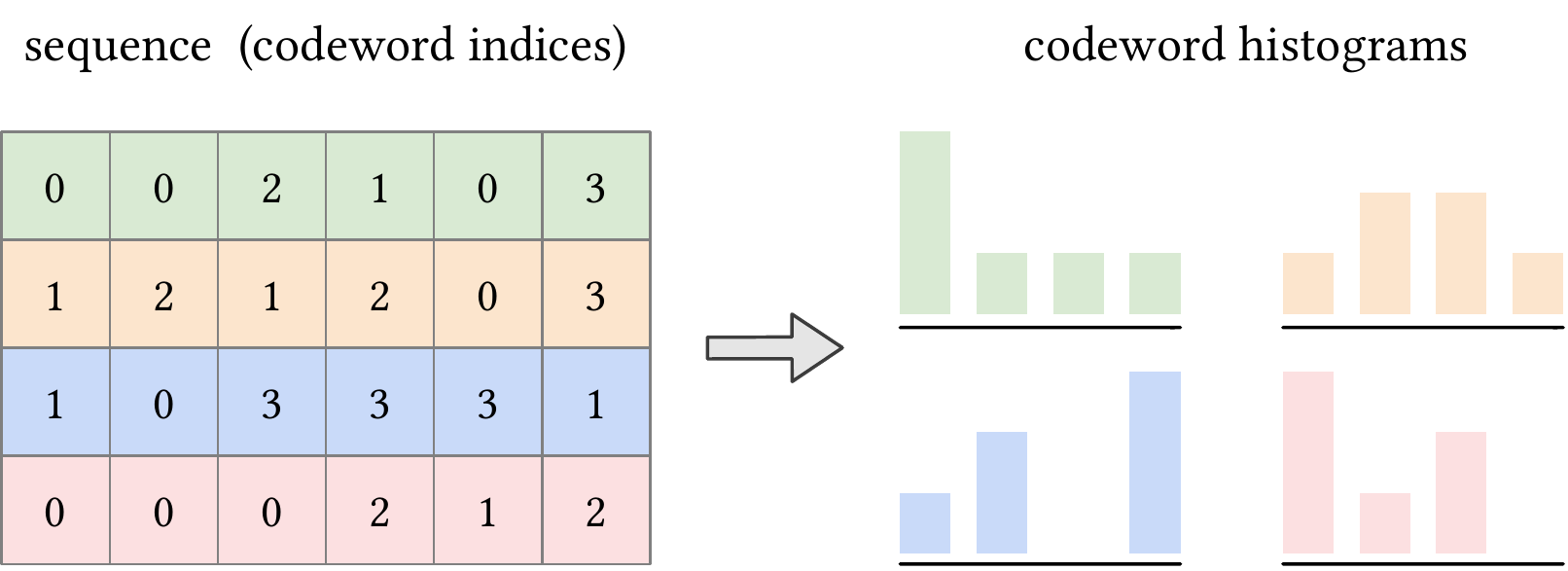}
	\caption{Illustration of codeword histograms. There are four codewords in each of the four codebooks.}
	\label{fig:codeword_histograms}
\end{figure}

Inspired by the benefit that redundant inner product computations can be circumvented in MIPS algorithms based on vector quantization, and the ability of using codebooks to quantize any embedding matrix, we propose LISA (LInear-time Self-Attention), an efficient attention mechanism based on computing codeword histograms. Equipped with a series of codebooks to encode items (or any form of tokens), LISA can dramatically reduce the costs of inner product computation in a similar vein. Since each item (token) is represented as a composition of codewords, and the entire input sequence can be compressed to a histogram of codewords for each codebook (illustrated in Figure~\ref{fig:codeword_histograms}), we are essentially performing attention over codewords. The histograms are used to compute the attention weights matrix in $O(L)$ time. We then pool over the codewords with the attention weights to get the outputs. To enable self-attention in a unidirectional setting (i.e., with casual masking~\cite{kitaev2019reformer}), we can resort to the mechanism of prefix-sums and compute a histogram at each position of the sequence. 

Compared to the efficient attention methods that rely on sparse patterns, our proposed method performs full contextual attention over the input sequence, with a computational and memory complexity linear in the sequence length. Our proposed method also enjoys the compression of item embeddings brought by LightRec. Particularly, in an online recommendation setting, our method can encapsulate a user's entire history with a fixed size histogram, greatly reduce the storage costs.

Our contributions can be summarized as follows:
\begin{itemize}[leftmargin=*]
    \item We propose LISA (LInear-time Self-Attention), a novel attention mechanism for efficient recommendation that reduces the complexity of computing attention scores from $O(L^2D)$ to $O(LBW)$, while simultaneously enabling model compression. The total number of codewords $BW$ is a budget hyperparameter balancing between performance and speed.
    \item We also propose two variants of LISA, one of them allows soft codeword assignments, and the other uses a separate codebook to encode sequences. These techniques allow us to use much smaller codebooks, resulting in further efficiency improvements.
    \item We conduct extensive experiments on four real-world datasets. Our proposed method obtains similar performance to vanilla self-attention, while significantly outperforms the state-of-the-art efficient attention baselines in both performance and efficiency.
\end{itemize}
\section{Related Work}

\subsection{Applications of Self-Attention Mechanisms}
The scaled dot product self-attention introduced in Transformers~\cite{vaswani2017attention} has been extensively used in natural language understanding~\cite{devlin2019bert,xu2020layoutlm}. As a powerful mechanism that connects all tokens in the inputs with a pooling operation based on relevance, self-attention has also made tremendous impacts in various other domains like computer vision~\cite{xiang2020learning,zhang2019self}, graph learning~\cite{velivckovic2017graph}. 

Recently, self-attention networks are successfully applied to sequential recommendation. \citeauthor{kang2018self}~\cite{kang2018self} adapted a Transformer architecture by optimizing the binary cross-entropy loss based on inner product preference scores, while \citeauthor{zhang2018next}~\cite{zhang2018next} propose to optimize a triplet margin loss based on Euclidean distance preference. Self-attention is also used for geographical modeling in location recommendation~\cite{lian2020geography,lian2020personalized}. They have demonstrated significant performance improvements over the RNN based models. 

\subsection{Improving Efficiency of Attention}
Considerable efforts have been made trying to scale Transformers to long sequences. Transformer-XL in~\cite{dai2019transformer} captures longer-term dependency by employing a segment-level recurrent mechanism, which splits the inputs into segments to perform attention. \citeauthor{sukhbaatar2019adaptive}~\cite{sukhbaatar2019adaptive} limited the self-attention context to the closest samples. However, these techniques do not improve the $O(L^2)$ asymptotic complexity of self-attention.

In another line of work, attempts in reducing the asymptotic complexity are made. \citeauthor{child2019generating}~\cite{child2019generating} proposed to factorize the attention computation into local and strided ones. \citeauthor{tay2020sparse}~\cite{tay2020sparse}, on the other hand, improved local attention by introducing a differentiable sorting network to re-sort the buckets. Reformer~\cite{kitaev2019reformer} hashes the query-keys into buckets via hashing functions based on random projection, and attention is computed within each bucket. In a similar manner, \citeauthor{roy2020efficient}~\cite{roy2020efficient} assign tokens to buckets through clustering. Built on top of ETC~\cite{ainslie2020encoding}, Big Bird~\cite{zaheer2020big} considers a mixture of various sparse patterns, including sliding window attention and random attention. Clustered Attention, introduced in~\cite{vyas2020fast}, however, groups queries into clusters and perform attention on centroids. Linformer~\cite{wang2020linformer} resorts to a low-rank projection on the length dimension. However, it can only operate in a bidirectional mode without casual masking. 

Most of the aforementioned approaches rely on sparse attention patterns, while our method performs full contextual attention over the whole sequence. Besides, Linformer and Sinkhorn Transformer assume a fixed sequence length due to the use of sorting network and projection, while our method poses no such constraint. Our method is also notably faster than the existing approaches, enjoying an asymptotic complexity of $O(L)$, while inner product can be stored in a table.
\section{Methodology}
In this section, we first quickly go through some of the underlying preliminaries. Then we introduce our proposed method step by step, starting from a simple case. We propose two more variants for further efficiency improvement. Finally, we analysis the complexity of our method.
\subsection{Preliminaries}
\subsubsection{Regular Self-Attention Mechanism}
The vanilla dot-product attention, introduced in \cite{vaswani2017attention}, accepts matrices $Q, K, V$ representing queries, keys and values, and computes the following outputs:
\begin{equation}
    V' = \operatorname{softmax} \left(\frac{QK^T}{\sqrt{D}}\right)V
    \label{eq:vanilla_attention}
\end{equation}
In the self-attention setting, we let the input sequence attend to itself. Concretely, given an input sequence $X \in \mathbb{R}^{L\times D}$, we linearly project $X$ via three matrices to get $Q=XP^Q, K=XP^K$ and $V=XP^V$. The results are then computed using Eq.~\eqref{eq:vanilla_attention}. This operation can be interpreted as computing a weighted average of the all other positions for every position in the sequence.

Self-attention has already been widely used in recommendation~\cite{kang2018self,sun2019bert4rec,xu2019graph,yu2019nairs}. \citeauthor{kang2018self}~\cite{kang2018self} used self-attention along with the feed-forward network from~\cite{vaswani2017attention} to encode user's sequential behaviors, and recommend the next item by computing the inner product between the encoded representation and target items' embeddings.

However, the computation of Eq.~\eqref{eq:vanilla_attention} suffers from quadratic computational and memory complexities, as computing the attention scores (the softmax term) and performing the weighted average both require $O(L^2 D)$ operations.

\subsubsection{Embedding Quantization with Codebooks}
Our efficient attention method is motivated by the idea of using codebooks to compress the embedding matrix~\cite{lian2020lightrec,jegou2010product,ge2013optimized,chen2018learning}. LightRec, proposed in~\cite{lian2020lightrec}, encodes items with a set of $B$ codebooks, each contains $W$ $D$-dimensional codewords that serve as a basis of the latent space. An item's embedding $x_i$ can be approximately encoded as:
\begin{equation}
    x_i \approx \sum_{b=1}^{B} c_{w_{i}^{b}}^{b}, \ \text {s.t.} w_{i}^{b}=\arg\max_{w} \operatorname{sim}(x_{i}, c_{w}^{b})
    \label{eq:codebook_encoding}
\end{equation}
where $\operatorname{sim} (x, y)$ is a similarity metric between two vectors $x,y$. In LightRec, a bilinear similarity function is adopted: $\operatorname{sim} (x, y)=x^T \boldsymbol{W} y + \left< \boldsymbol{w_1}, x\right> + \left< \boldsymbol{w_2}, y \right> $. $c^b_w$ denotes the $w$-th codeword in the $b$-th codebook. $\boldsymbol{W}, \boldsymbol{w_1}, \boldsymbol{w_2}$ are learnable weights. 

At training time, the codebooks and the item embeddings can be jointly trained using a softmax relaxation and the straight-through estimator~\cite{bengio2013estimating}. At the inference stage, the item embedding $x_i$ can be discarded completely. For each item $i$, we only store its corresponding codeword indices in each codebook, i.e., $[w^1_i, w^2_i,\dots,w^B_i]\in \{1, \dots, W\}^B $. Because each codeword index can be encoded with $\log_2 W$ bits, the memory cost of storing $N$ items is reduced from $4ND$ bytes to $\frac{1}{8}NB\log_2 W + 4BWD$ bytes, where the first term is for codeword indices, and the second term is for codebooks.

\subsection{Motivation: A Simple Case}
\begin{figure}
	\centering
	\includegraphics[width=.95\linewidth]{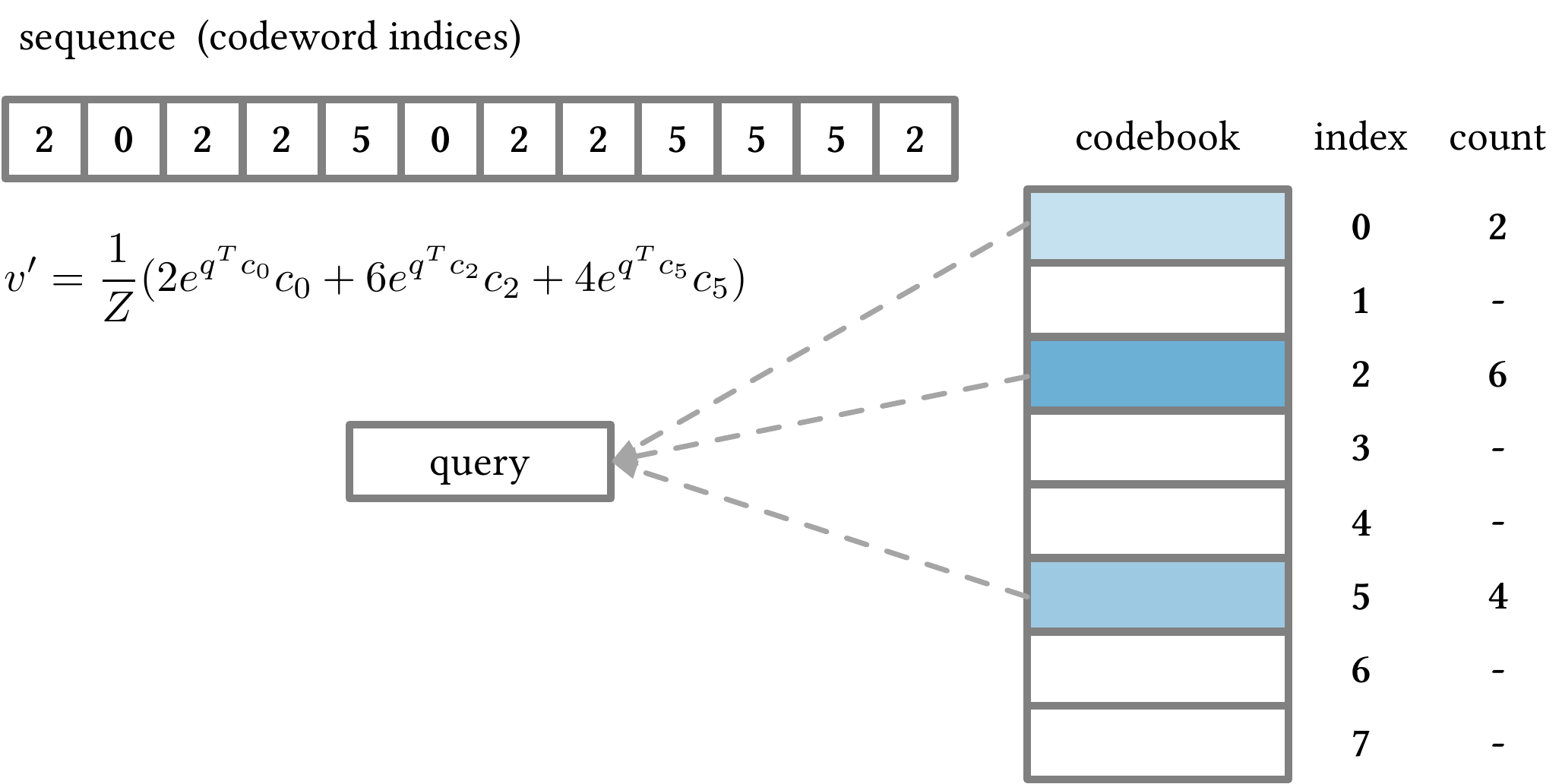}
	\caption{Example of using codeword histogram to avoid redundant computation in attention, where a single codebook is used. Here $v'$ is the attention output for a given query $q$, $Z$ is a normalization constant. }
	\label{fig:single_codebook_example}
\end{figure}

To illustrate the motivation behind our proposed method, we first look at a simple case where a single codebook is used to encode items. 

In this case, an item is directly represented by the codeword with the maximum relevance score to it. The $i$-th item in the sequence $X$ is therefore given by: $x_i=c_{w_i}$, where $w_i=\arg\max_{w} \operatorname{sim} (x_i, c_w)$ and $c_w \in \mathbb{R}^D$ denotes the $w$-th codeword in the codebook. Then, to perform the dot-product attention for a query $q\in \mathbb{R}^D$ (with keys and values being the sequence $X$), we compute the inner product between $q$ and the corresponding codeword $c_{w_i}$ for every item in the sequence. The output $v'$ of the attention is computed as follows: 
\begin{equation}
    v' = \frac{\sum_{i=1}^L \exp{(q^T x_{i})} x_{i}}{\sum_{i=1}^L \exp(q^T x_{i})} = \frac{\sum_{i=1}^L \exp{(q^T c_{w_i})} c_{w_i}}{\sum_{i=1}^L \exp(q^T c_{w_i})} 
\label{eq:attn_single_codebook_orig}
\end{equation}
where $L$ is the sequence length. For the sake of simplicity, we omit the projection matrices $P^Q, P^K, P^V$ at this moment. From the above equation, we observe that we may have repeatedly compute the inner product of $q$ with the same codeword $c_w$, since a number of items in the sequence may all share $c_w$ as their representations. This redundant computation significantly hampers the efficiency, especially when $L \gg \lvert \Omega_X \rvert$, where $\Omega_X$ is the set of unique codeword indices that the $L$ items in the sequence correspond to, i.e., $\Omega_X=\{w \mid \exists i: w=w_i\}$.

To address this issue, we note that $v'$ is just a weighted average of all the codewords in $\Omega_X$, and the weight of each codeword depends only on its inner product with $q$ and its number of occurrences. Therefore, we only need to count how many times each codeword $c_w$ in $\Omega_X$ is used in the sequence, and compute the inner product of $q$ with $c_w$ once. The computation of Eq.~\eqref{eq:attn_single_codebook_orig} can be reformulated as:
\begin{equation}
    v' = \frac{\sum_{w \in \Omega_X} f_w \exp{(q^T c^b_w)} c^b_w}{\sum_{w \in \Omega_X} f_w \exp{(q^T c^b_w)}} 
\end{equation}
We illustrate this idea in Figure~\ref{fig:single_codebook_example}.

\subsection{Linear-Time Self Attention}
As we can see, the mechanism of codebook allows us to obtain the exact results of dot-product attention with less computation (both in computing the attention scores and computing the final weighted average), at least in the case of a single codebook. Now, we turn to the case that multiple codebooks are used. The items in the sequence are represented by an additive composition of codewords in all codebooks, as given by Eq.~\ref{eq:codebook_encoding}. The result of dot-product attention for a given query $q$ is as follows:
\begin{equation}
    v' = \frac{\sum_{i=1}^L \exp{(q^T \sum_{b=1}^B c_{w^b_i})} \sum_{b=1}^B c_{w^b_i}}{\sum_{i=1}^L \exp(q^T \sum_{b=1}^B c_{w^b_i})} 
\end{equation}
Unlike the single codebook scenario, although in each codebook many items may correspond to the same codeword, their representations will diverge after the additive composition. Hence we still have to compute the inner product between $q$ and every item $x_i$ in the sequence.

\begin{figure*}
	\centering
	\includegraphics[width=.9\linewidth]{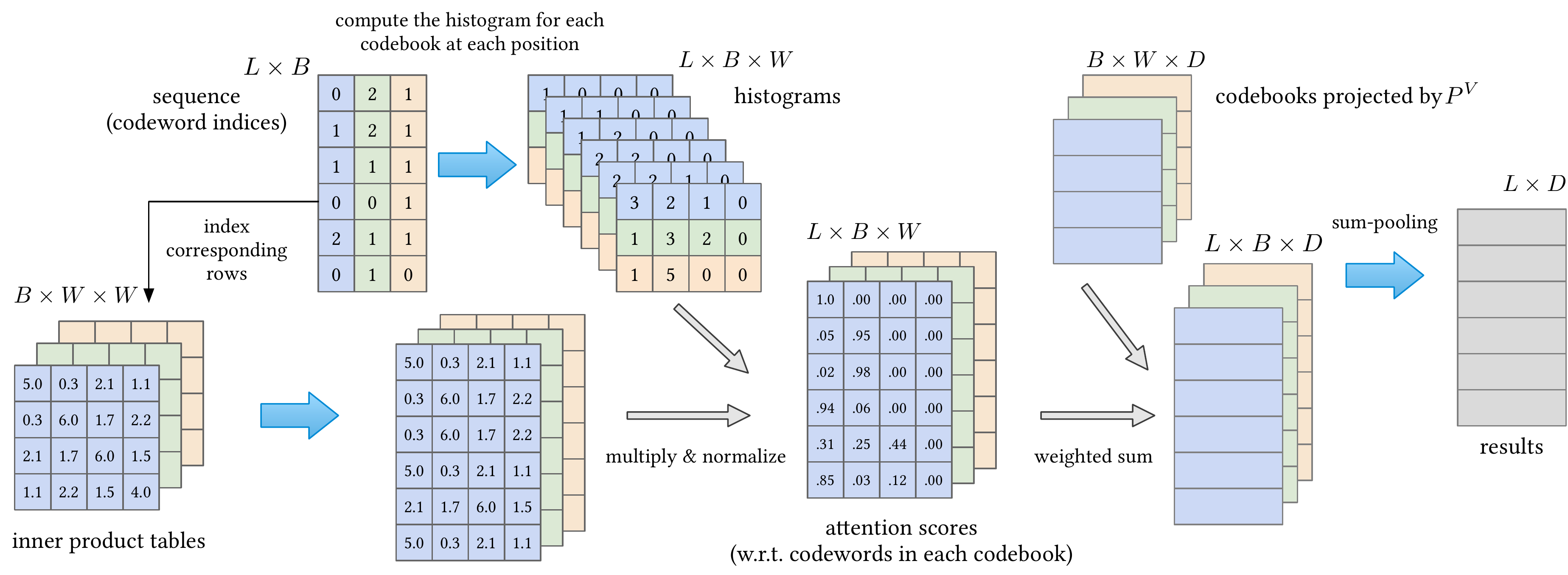}
	\caption{Workflow of LISA (unidirectional mode). Codeword indices of the items in the sequence are taken as input, they are used to index the inner product tables, and compute the codeword histograms for each codebook at every position. The histograms are element-wise multiplied with the inner product extracted from the table, and then normalized to obtain the attention scores, which are used to compute the weighted average of the projected codebooks. In this example, three codebooks are used, each contains four codewords.}
	\label{fig:workflow}
\end{figure*}

To tackle this problem, we propose to relax the attention operation. We split the computation, perform the attention in each codebook separately, and then take the sum:
\begin{equation}
    v'' = \sum_{b=1}^B \frac{\sum_{i=1}^L \exp{(q^T c^b_{w^b_i})} c^b_{w^b_i}}{\sum_{i=1}^L \exp{(q^T c^b_{w^b_i}})}
    \label{eq:attn_multi_codebooks_relax}
\end{equation}
This additive compositional formulation can be considered as a form of "multi-head" attention, where each attention head correlates with a codebook. Since different codebooks form different latent spaces, Eq.~\eqref{eq:attn_multi_codebooks_relax} in fact, aggregates information from different representational
subspaces of the items, using independent attention weights. 

Equipped with the above relaxation, we can once again reuse the inner product by computing the frequencies of each codeword that appeared in the sequence, separately for every codebook. We can reformulate the computation of Eq.~\eqref{eq:attn_multi_codebooks_relax} as follows:
\begin{equation}
    v'' = \sum_{b=1}^B \frac{\sum_{w\in \Omega^b_X} f^b_w \exp{(q^Tc^b_w)} c^b_w}{\sum_{w\in \Omega^b_X} f^b_w \exp{(q^Tc^b_w)}}
    \label{eq:attn_multi_codebooks_relax_count}
\end{equation}
where $\Omega^b_X=\{w \mid \exists i: w=w^b_i\}$ is the set of unique codeword indices of the $b$-th codebook, and $f^b_w$ is the number of occurrences of $c^b_w$ in the sequence.

However, the cardinality of $\Omega^b_X$ varies across different sequences $X$ and different codebooks $b$. The computation of Eq.~\eqref{eq:attn_multi_codebooks_relax_count} therefore operates on different sizes of tensors, which is sub-optimal for efficient batching in GPU and TPU~\cite{kochura2019batch}.
For batching purpose, we perform the attention over all codewords in each codebook, fixing the "context size" of the attention to $W$:
\begin{equation}
    v'' = \sum_{b=1}^B \frac{\sum_{w=1}^W f^b_w \exp{(q^Tc^b_w)} c^b_w}{\sum_{w=1}^W f^b_w \exp{(q^Tc^b_w)}}
    \label{eq:attn_multi_codebooks_batching}
\end{equation}
For a codeword $c^b_w$ that is not used by any item in the sequence, the occurrence count $f^b_w=0$, $c^b_w$ will not contribute to the weighted average. Hence Eq.~\eqref{eq:attn_multi_codebooks_relax_count} and Eq.~\eqref{eq:attn_multi_codebooks_batching} is equivalent. 

Now we put it to the self-attention setting, where we use the input sequence as queries to attend to itself, the $i$-th query is just $x_i$, i.e., $q_i=x_i=\sum_{b=1}^B c^b_{w^b_i}$. Since we regard the attention in different codebook as independent heads that attend in different latent spaces, we further reduce the computation of the inner product from $q_i^Tc^{b'}_{w'}=\sum_{b=1}^B c^b_{w^b_i} c^{b'}_{w'}$ to $c^{b'}_{w^{b'}_i} c^{b'}_{w'}$, considering only the term in the same codebook. This gives us:
\begin{equation}
    x'_i = \sum_{b=1}^B \frac{\sum_{w'=1}^W f^b_{w'} \exp{(c^{b^T}_{w^b_i}c^b_{w'})} c^b_{w'}}{\sum_{w'=1}^W f^b_{w'} \exp{(c^{b^T}_{w^b_i} c^b_{w'})}}
    \label{eq:self_attn_multi_codebooks}
\end{equation}
where $x'_i$ is the $i$-th output of the attention operation. 

Eq.~\eqref{eq:self_attn_multi_codebooks} computes the bidirectional attention (each position can attend over all positions in the input sequence), since $f^b_{w'}$ indicates the frequency of $c^b_{w'}$ in the entire sequence. However, in the recommendation setting, the model should consider only the first $i$ items when making the $i+1$-th prediction~\cite{kang2018self}, we therefore favor a unidirectional setting (each position can only attend to positions up to and including that position). This requires us to compute the codeword histogram of every codebook, up to $i$-th position, for each $i=1, \dots, L$. This can be implemented via the mechanism of prefix-sum. We first transform the codeword index $w^b_i$ into a one-hot representation $e^b_i = \operatorname{one-hot}(w^b_i)$, where $e^b_i \in \{0, 1\}^{W}$. The one-hot vectors $e^b_i$ for each codebook $b$ at each position $i$ forms a tensor $\mathbf{E}$ of shape $L\times B\times W$, we compute the prefix-sum along the first dimension to get the histograms up to each position in the sequence:
\begin{equation}
    \mathbf{F}_{i,:,:} = \sum_{j=1}^i \mathbf{E}_{i,:,:}
    \label{eq:prefix_sum}
\end{equation}

There exists an efficient algorithm~\cite{cormen2009introduction,ladner1980parallel} for prefix-sum with a computational complexity of $O(\log L)$ when compute in parallel.

As we mentioned earlier, in the vanilla self-attention~\cite{vaswani2017attention}, linear projections are applied on the input sequence $X$ to get queries, keys and values. Similarly, we can directly apply the projection matrices $P^Q, P^K, P^V$ on the codebooks since every item in the input sequence $X$ is just a composition of codewords. Combining this with Eq.~\eqref{eq:prefix_sum}, we obtain the following unidirectional attention mechanism:
\begin{equation}
    x'_i = \sum_{b=1}^B \frac{\sum_{w'=1}^W \mathbf{F}_{i,b,w'} \exp{\left(({P^Qc^{b}_{w^b_i})}^T(P^Kc^b_{w'})\right)} P^Vc^b_{w'}}{\sum_{w'=1}^W \mathbf{F}_{i,b,w'} \exp{\left({(P^Qc^{b}_{w^b_i})}^T(P^Kc^b_{w'})\right)}}
\end{equation}

\begin{algorithm}[tb]
\caption{LISA}
\SetKwInOut{Input}{Input}
\SetKwInOut{Output}{Output}
\SetKwInOut{Parameters}{Parameters}
\label{alg:lisa_unidirectional}
\Input{Codeword indices of the sequence $\Psi \in \mathbb{N}^{L\times B}$}
\Parameters{Inner product table (after taking the exponent) $\mathbf{M} \in \mathbb{R}^{B\times W\times W}$, where $\mathbf{M}_{b,i,j}=\exp{\left((P^Q c^b_{i})^T (P^K c^b_{j})\right)}$; projected codebooks for values $\mathbf{C}^V=\mathbf{C}P^V \in \mathbb{R}^{B\times W \times D}$}
\Output{The results of self-attention $X' \in \mathbb{R}^{L\times D}$}
Convert $\Psi$ to one-hot representations $\mathbf{E}=\operatorname{one-hot}(\Psi) \in \{0,1\}^{L\times B\times W}$\;

\eIf{unidirectional mode}{
Compute the prefix-sums $\mathbf{F}^{L\times B \times W}$ of $\mathbf{E}$ along the first dimension according to Eq.~\eqref{eq:prefix_sum}\;}{
Compute the sums $\mathbf{F}^{B\times W}$ of $\mathbf{E}$ along the first dimension and broadcast $\mathbf{F}$ to shape $L\times B\times W$\;}

Gather inner product $\mathbf{S} \in \mathbb{R}^{L\times B \times W}$ from $\mathbf{M}$ along the first two dimensions using indices $\Psi$\;
$\mathbf{A} = \mathbf{F} \otimes  \mathbf{S}$ (element-wise multiplication)\;
Normalize the attention scores $A$ along the last dimension\;
$X'= \sum_{b=1}^B \mathbf{A}_{:,b,:}\mathbf{C}^V_{b,:,:}$\;
\Return X'
\end{algorithm}

As we only need to compute the inner product between codewords in the same codebook, we can store them (after taking the exponent) in tables, and retrieve required terms via table lookup at inference time. We can achieve this by storing $B$ tables with $W^2$ items each, resulting in a memory cost of $O(BW^2)$. However, this is not feasible without embedding quantization via codebooks, which leads to memory complexity of $O(N^2)$, where $N \gg BW$ is the number of items. We present the workflow of LISA in Figure~\ref{fig:workflow}, and we outline the main algorithm for LISA formally in Algorithm~\ref{alg:lisa_unidirectional}.

\subsection{Variants}
We notice that the computational cost of LISA is determined by the fixed context size of $B\times W$ (i.e., the total number of codewords). To further increase the efficiency, especially on shorter sequences, we propose to use a separate set of codebooks to encode the sequence with a much smaller $B\times W$. In our experiments, we find that using a $B\times W$ of 128/256 is enough to obtain decent performance, compared to a $B \times W$ of 1024/2048 that we used in our base model. We investigate the following two variants:
\begin{itemize}[leftmargin=*]
    \item \textbf{LISA-Soft}: Instead of assigning a unique codeword $w^b_i$ for each item $i$, we allow a soft codeword assignment. In this case, $\mathbf{E}$ becomes the softmax scores where $\mathbf{E}_{i,b,:}=\operatorname{softmax}(\operatorname{sim}(x_i, C^b))$. With a soft assignment we can no longer compress the embedding matrix by storing discrete codeword indices at inference time. Hence we directly use the original embeddings for target items.
    \item \textbf{LISA-Mini}: To enable embedding compression, we still use a hard codeword assignment. We adopt two separate sets of codebooks: a smaller one (i.e., with a smaller $B\times W$) to encode the sequence, and a larger one to encode target items.
\end{itemize}

\subsubsection{Extensions}
Vanilla Transformer could stack multiple self-attention layers to improve performance. However, we find that using multiple attention layers is not particularly helpful in recommendation, as with~\cite{kang2018self}. Therefore we only employ a single layer. Our method can easily extend to multiple layer cases. A straightforward solution is to use a different set of codebooks to remap the attention outputs to codewords in a different set of latent spaces. Our method can also be adapted beyond self-attention, as long as queries, keys and values can be encoded via codebooks. Besides recommendation, we can employ LISA in other domains since codebooks are able to quantize any embedding matrices. For example, the inputs in NLP tasks are just token embeddings, where our method can easily be applied. 

\subsection{Complexity Analysis}
We see that computing the codeword histograms takes $O(LBW)$ steps, as we have to compute the prefix-sums along the sequence length dimension for every codeword in all codebooks. The time complexity for computing the final outputs (weighted sum of values) is $O(LBWD)$, as this operation is essentially a batched matrix multiplication between an attention score tensor of shape $L\times B\times W$ and a tensor of shape $B\times W \times D$ representing codebooks. Computing the inner product tables requires $O(BW^2D)$ time, but at inference time, we can save this cost via table lookup. At training time, this is still a negligible term compared to $O(NLBWD)$, where $N$ is the batch size. Hence, our method has an overall asymptotic time complexity of $O(LBWD)$.
\section{Experiments}
In this section, we empirically analyze the recommendation performance of our proposed method, compared to the vanilla Transformer and existing efficient attention methods. Following that, we present the computational and memory costs of LISA with respect to different sequence lengths. We also investigate how the number of codewords affects the performance of our method. Finally, we show the efficiency improvement brought by LISA in an online setting. We have also published our code\footnote{Available at: \url{https://github.com/libertyeagle/LISA}}.
 
\begin{table}
    \caption{Dataset statistics.}
    \label{tab:datasets}
    \centering
    \begin{tabular}{lrrrr}
        \toprule
        Dataset & \#users & \#items & \#ratings & avg. length \\
        \midrule
        Alibaba & \numprint{99979} & \numprint{80000} & 25M & 252.93 \\
        ML-1M & \numprint{6040} & \numprint{3416} & 1M & 165.50 \\
        Video Games & \numprint{59766} & \numprint{33487} & 0.5M & 8.82 \\
        ML-25M & \numprint{162541} & \numprint{32720} & 25M & 153.47 \\
        \bottomrule
    \end{tabular}
\end{table}

\subsection{Datasets}
We use four real-world datasets for sequential recommendation that vary in platforms, domains and sparsity:
\begin{itemize}[leftmargin=*]
    \item \textbf{Alibaba}: A dataset sampled from user click logs on Alibaba e-commerce platform, collected from September 2019 to September 2020. This is a dataset that contains relatively longer behavior sequences than the other datasets used in the experiments.
    \item Amazon \textbf{Video Games}~\cite{ni2019justifying}: A series of product reviews data crawled from Amazon spanning from 1996 to 2018. The data is split into separate datasets according to the top-level product categories. In this work, we consider the "Video Games" category. This dataset is notable for its sparsity.
    \item \textbf{MovieLens}~\cite{harper2016movielens}: A widely used benchmark dataset of movie ratings for evaluating recommendation algorithms. We adopt two versions: MovieLens 1M (\textbf{ML-1M}) and MovieLens 25M (\textbf{ML-25M}), which include 1 million and 25 million ratings, respectively.
\end{itemize}
Following the common pre-processing practice in \cite{kang2018self,sun2019bert4rec,tang2018personalized}, we treat the presence of a rating as implicit feedback. Users and items with fewer than five interactions are discarded. Table~\ref{tab:datasets} shows the statistics of the processed dataset.

\begin{table}[b]
    \caption{Settings for LISA and achieved compression ratio on each dataset (shown in the last four rows) . We present the settings of codebooks used to encode sequence. LISA-Mini also applies a separate set of codebooks to encode target items, with the same settings as LISA-Base.}
    \label{tab:settings_lisa}
    \centering
    \begin{tabular}{lccc}
        \toprule
        & LISA-Base & LISA-Soft & LISA-Mini\\
        \midrule
        \#codebooks ($B$) & 8 & 8 & 8 \\
        \midrule
        \#codewords ($W$) & \tabincell{c}{128 (ML-1M)\\256 (Others)} & 16 & 32   \\
        \midrule
        \textit{Alibaba} & 24.26 & - & 18.45 \\
        \midrule
        \textit{ML-1M} & 3.19 & - & 2.51 \\
        \midrule
        \textit{Video Games} & 13.02 & - & 10.62 \\
        \midrule
        \textit{ML-25M} & 12.78 & - & 10.44\\
        \bottomrule
    \end{tabular}
\end{table}

\subsection{Compared Methods}
We evaluate our proposed base model, denoted as \textbf{LISA-Base}, as well as its two variants: LISA-Soft and LISA-Mini. We compare these methods with the vanilla Transformer~\cite{vaswani2017attention}, as well as the following efficient attention methods:
\begin{itemize}[leftmargin=*]
    \item \textbf{Reformer}~\cite{kitaev2019reformer}: It utilizes LSH to restrict queries to only attend to keys that fall in the same hash bucket, reducing the computational complexity to $O(L\log L)$. We do not use the reversible layers since this technique can be applied to all methods, including ours.
    \item \textbf{Sinkhorn} Transformer~\cite{tay2020sparse}: It extends local attention by learning a differentiable sorting of buckets. Queries can then attend to keys in the corresponding sorted bucket. This model has a computational complexity of $O(LB + (\frac{L}{B})^2)$, where $B$ is the bucket size.
    \item \textbf{Routing} Transformer~\cite{roy2020efficient}: It is a clustering-based attention mechanism. K-means clustering is applied to input queries and keys. The attention context for a query is restricted to keys that got into the same cluster with the query. The computational complexity is $O(Lk+\frac{L^2}{k})$, where $k$ is the number of clusters.
    \item Improved \textbf{Clustered} Attention~\cite{vyas2020fast}: Another clustering-based attention method. This approach, however, only groups queries into clusters, and attend cluster centroids over all keys. The top-$k$ keys for each cluster centroid are extracted to compute the attention scores with queries in this cluster. This results in a computational complexity of $O(LC+Lk)$, where $C$ is the number of clusters.
    \item \textbf{Linformer}~\cite{wang2020linformer}: An efficient attention mechanism based on low-rank approximation. Linformer projects the keys and values of shape $L\times D$ to $k \times D$, effectively reducing the context size to a tunable hyperparameter $k$. This leads to a complexity of $O(Lk)$. We note that it is the only baseline that does not support unidirectional attention.
\end{itemize}
For simplicity, we ignore the term regarding the latent dimension size $D$ in the above-mentioned asymptotic complexities.

\begin{table*}
    \caption{Recommendation performance on Alibaba, ML-1M and Video Games. The number in the parentheses in baseline methods denotes the bucket size used for Sinkhorn Transformer and Routing Transformer, and \#clusters for Clustered Attention. Bold font denotes the best-performing method among the efficient attention baselines, LISA-Soft and LISA-Mini.}
    \label{tab:rec_results_part_1}
    \resizebox{0.98\textwidth}{!}{
    \begin{tabular}{lcccccccccccc}
        \toprule
        & \multicolumn{4}{c}{Alibaba} & \multicolumn{4}{c}{ML-1M} & \multicolumn{4}{c}{Video Games} \\
        \cmidrule(lr){2-5}
        \cmidrule(lr){6-9}
        \cmidrule(lr){10-13}
        & HR@5 & NDCG@5 & HR@10 & NDCG@10 & HR@5 & NDCG@5 & HR@10 & NDCG@10 & HR@5 & NDCG@5 & HR@10 & NDCG@10 \\
        \midrule
        Transformer & 0.6597 & 0.5528 & 0.7569 & 0.5843 & 0.6841 & 0.5376 & 0.7914 & 0.5725 & 0.5525 & 0.4337 & 0.6583 & 0.4680 \\
        \midrule
        Linformer & 0.3829 & 0.3007 & 0.4929 & 0.3360 & 0.4171 & 0.2899 & 0.5704 & 0.3394 & 0.4643 & 0.3605 & 0.5671 & 0.3937 \\
        Reformer (LSH-1) & 0.6209 & 0.5189 & 0.7212 & 0.5513 & 0.6753 & 0.5248 & 0.7806 & 0.5590 & 0.5637 & 0.4429 & 0.6694 & 0.4771 \\
        Reformer (LSH-4) & 0.6184 & 0.5156 & 0.7199 & 0.5484 &	0.6492 & 0.5040 & 0.7627 & 0.5408 & 0.5648 & 0.4446 & 0.6685 & 0.4781 \\
        Sinkhorn (32) & 0.6298 & 0.5278 & 0.7260 & 0.5589 & 0.6743 & 0.5256 & 0.7796 & 0.5599 & 0.5479 & 0.4289 & 0.6557 & 0.4638 \\
        Sinkhorn (64) & 0.6331 & 0.5319 & 0.7289 & 0.5629 & 0.6775 & \textbf{0.5310} & 0.7844 & \textbf{0.5656} & 0.5469 & 0.4258 & 0.6541 & 0.4605 \\
        Routing (32) & 0.5742 & 0.4789 & 0.6724 & 0.5106 & 0.6623 & 0.5186 & 0.7704 & 0.5537 & 0.5615 & 0.4412 & 0.6657 & 0.4750 \\
        Routing (64) & 0.6037 & 0.5037 & 0.7023 & 0.5356 & 0.6535 & 0.5100 & 0.7616 & 0.5452 & 0.5570 & 0.4369 & 0.6604 & 0.4704 \\
        Clustered (100) & 0.5924 & 0.4937 & 0.6941 & 0.5266 & 0.6573 & 0.5127 & 0.7697 & 0.5492 & 0.5591 & 0.4394 & 0.6642 & 0.4734 \\
        Clustered (200) & 0.5934 & 0.4936 & 0.6962 & 0.5268 & 0.6538 & 0.5095 & 0.7712 & 0.5478 & 0.5578 & 0.4384 & 0.6633 & 0.4725 \\
        \midrule 
        LISA-Base & 0.6660 & 0.5460 & 0.7702 & 0.5798 & 0.6940 & 0.5406 & 0.7962 & 0.5740 & 0.6203 & 0.4788 & 0.7338 & 0.5157 \\
        \midrule
        LISA-Soft & \textbf{0.6575} & \textbf{0.5393} & \textbf{0.7622} & \textbf{0.5732} & 0.6795 & 0.5229 & \textbf{0.7887} & 0.5587 & \textbf{0.5951} & \textbf{0.4592} & 0.7035 & \textbf{0.4944} \\
        LISA-Mini & 0.6430 & 0.5146 & 0.7559 & 0.5511 & \textbf{0.6853} & 0.5308 & 0.7886 & 0.5644 & 0.5917 & 0.4497 & \textbf{0.7102} & 0.4881 \\
        \bottomrule
    \end{tabular}
    }
\end{table*}

\begin{table}
    \caption{Recommendation performance on ML-25M.}
    \label{tab:rec_results_part_2}
    \resizebox{0.98\columnwidth}{!}{
    \begin{tabular}{lcccc}
        \toprule
        & HR@5 & NDCG@5 & HR@10 & NDCG@10 \\
        \midrule
        Transformer & 0.9338 & 0.8073 & 0.9752 & 0.8209\\
        \midrule
        Linformer & 0.8627 & 0.7086 & 0.9367 & 0.7329\\
        Reformer (LSH-1) & 0.9214 & 0.7847 & 0.9694 & 0.8005 \\
        Reformer (LSH-4) & 0.9150 & 0.7765 & 0.9667 & 0.7935  \\
        Sinkhorn (32) & 0.9195 & 0.7836 & 0.9682 & 0.7995\\
        Sinkhorn (64) & 0.9161 & 0.7820 & 0.9649 & 0.7980 \\
        Routing (32) & 0.9167 & 0.7829 & 0.9658 & 0.7990 \\
        Routing (64) & 0.9215 & 0.7890 & 0.9685 & 0.8044 \\
        Clustered (100) & 0.9215 & 0.7830 & 0.9700 & 0.7989\\
        Clustered (200) & 0.9199 & 0.7818 & 0.9692 & 0.7980\\
        \midrule 
        LISA-Base & 0.9254 & 0.7933 & 0.9713 & 0.8083 \\
        \midrule
        LISA-Soft & \textbf{0.9269}	& \textbf{0.7964} & \textbf{0.9710} & \textbf{0.8109}\\
        LISA-Mini & 0.9243 & 0.7900 & 0.9701 & 0.8050\\
        \bottomrule
    \end{tabular}
    }
\end{table}

\subsection{Settings \& Metrics}
\subsubsection{Parameter Settings}
We use the SASRec~\cite{kang2018self} architecture as the building block for our experiment setup, as SASRec purely relies on self-attention to perform sequential recommendation. Hence we can simply replace the regular Transformer self-attention with our method or the aforementioned baselines to compare the performance. We find that the number of attention layers has negligible impacts on the recommendation performance, and the performance of using multiple attention heads is consistently worse than single head~\cite{kang2018self}. Multiple attention layers and attention heads only lead to greater computational cost. Hence we use a single layer and a single head for all compared methods. 

All methods are implemented in PyTorch and trained with the Adam optimizer with a learning rate of 0.001 and a batch size of 128. We use an embedding dimension of 128, and the dropout rate is set to 0.1 on all datasets. We train all methods for a maximum of 200 epochs. Following the settings in the original papers, we consider two settings for Reformer: LSH-1 and LSH-4, which use one and four parallel hashes, respectively. For Sinkhorn Transformer and Routing Transformer, we consider a bucket (window) size of 32 and 64. We set the number of clusters to 100 and 200 for Clustered Attention. We use a low-rank projection size of 128 for Linformer. We apply casual masking for all methods except Linformer.

We report the settings of codebooks used for all three versions of our proposed method in Table~\ref{tab:settings_lisa}. Since LISA-Base and LISA-Mini can simultaneously compress the embedding matrix, we also report the achieved compression ratios on all four datasets. We see that the item embeddings can be compressed up to 24x.

\subsubsection{Metrics}
Following~\cite{kang2018self,sun2019bert4rec,lian2020geography}, we apply two widely used metrics of ranking evaluation: Hit Rate (HR) and NDCG~\cite{weimer2008cofi}. HR@$k$, counts the fraction of times that the target item is among the top-$k$. NDCG@$k$, rewards methods that rank positive items in the first few positions of the top-$k$ ranking list. We report the two metrics at $k=5$ and $k=10$. The last item of each user's behavior sequence is used for evaluation, while the remaining are used for training. For each user, We randomly generate 100 negative samples that the user has not interacted with, pairing them with the positive sample for the compared methods to rank.

\subsection{Recommendation Performance}
\label{sec:rec_perf}
We report the results of the comparison of recommendation performance with baselines in Table~\ref{tab:rec_results_part_1} and Table~\ref{tab:rec_results_part_2}. Since we also care about efficiency besides performance, we use bold font to denote the best-performing method among the efficient attention baselines and the two more efficient variants of our approach, excluding LISA-Base.

\begin{figure*}
	\centering
	\includegraphics[width=.85\linewidth]{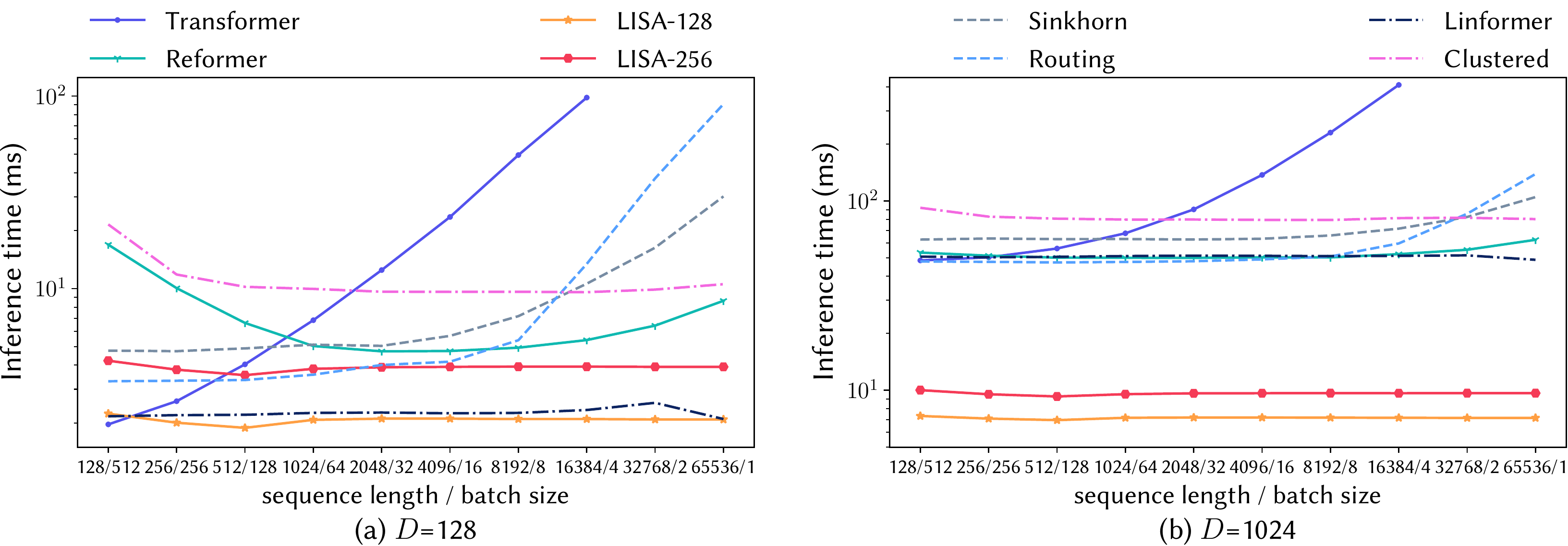}
	\caption{Inference speed of different methods. Note that the y-axis is in a logarithmic scale. For Transformer the plot is shown up to 16K sequences, as longer sequences will produce an out of memory error on a 16GB V100 GPU.}
	\label{fig:inference_speed}
\end{figure*}

From the two tables, we have the following important findings:
\begin{itemize}[leftmargin=*]
    \item \textit{LISA-Base consistently outperforms all the state-of-the-art efficient attention baseline on all four datasets}. It attains improvements of up to 8.78\% and 7.29\% over the best-performing baseline in terms of HR@10 and NDCG@10. This demonstrates the effectiveness of our proposed attention method based on codeword histograms, as we compute the full contextual attention, compared to the sparse attention mechanism most baselines built upon. Since we use more codewords, LISA-Base also outperforms LISA-Soft and LISA-Mini on all datasets except ML-25M, where it has similar performance to LISA-Soft. On some metrics and datasets, LISA-Base even obtains higher performance than Transformer. This does make sense, considering LISA-Base could attend over a broader context, encapsulating relevant information from a large number codewords in each codebook (providing diverse views).
    \item \textit{LISA-Soft and LISA-Mini achieve decent performance with much smaller codebooks.} Even with 16 codewords used per codebook, LISA-Soft still outperforms the best-performing baseline by 2.46\% and 1.16\% in terms of HR@10 and NDCG@10, on average. Only on ML-1M, it is slightly worse than Sinkhorn Transformer (64) in terms of NDCG. We suppose that the issue might be we only use the soft codeword assignment scores when computing the codeword histograms $\mathbf{F}$, we still use a unique codeword $c^b_{w^b_i}$ per codebook to approximate the query. Otherwise, it would pose challenges to handle the cross terms between different codebooks when computing the inner product. This could create a potential mismatch between queries and keys/values, leading to the performance gap on this dataset. However, in most cases, LISA-Soft achieves comparable performance with respect to LISA-Base, using 94\% fewer codewords. Even when model compression is desired, LISA-Mini can still improve the best-performing baseline by 2.46\% in terms of HR@10, on average.
    \item \textit{Our proposed method, and the ones that allocate items to buckets based on similarity, even lead to increased recommendation performance on Video Games dataset.} With an average length of only 8.8, the user sequences in Video Games tend to be noisy for making next-item recommendations. Full-context attention in this scenario would confuse the model with the noise. Reformer, Routing Transformer and Clustered Attention remedy this issue by only attending to the informative items selected through hashing or clustering (note that the number of buckets/clusters are predetermined according to the maximum sequence length in the dataset and the desired bucket/cluster size). Meanwhile, LISA addresses this issue by summarizing information from different codebooks, which can be reckoned as a way of denoising. 
    \item \textit{In general, sparse attention via sorting the buckets seems to be more effective than learning the bucket assignments.} We observe that Sinkhorn Transformer is a strong baseline, considerably outperforms Reformer, Routing Transformer and Clustered Attention on Alibaba and ML-1M, while has almost identical performance with them on ML-25M. Only on Video Games it performs slightly worse, due to the above-mentioned intrinsic noise in this dataset. In this instance, Sinkhorn Transformer will perform full contextual attention, as it divides the sequence into consecutive blocks of fixed size. 
    \item \textit{LSH is better than clustering in bucket assignment.} Reformer and Routing Transformer are both content-based sparse attention methods that differ mostly by the technique used to infer sparsity patterns. Reformer employs LSH while Routing Transformer resorts to online k-means clustering. We see that Reformer consistently outperforms Routing Transformer. The latter one sorts tokens by their distances to each cluster centroid and assigns membership via the top-k threshold. The centroids are updated by an exponential moving average of training examples. Unlike Reformer, this approach does not guarantee that each item belongs to a single cluster, which may partially contribute to Routing Transformer's worse performance.
    \item \textit{Unidirectional attention is vital for satisfactory performance in recommendation.} Observing the results of Linformer we can obtain this conclusion. Because the projection is applied to the length dimension, causing the mixing of sequence information, it is nontrivial to apply casual masking for Linformer. This bidirectional attention leads to significant performance degradation, as our attempts with other methods in bidirectional mode corroborate this finding. The designs of certain baseline methods also induce some issues in order to enforce casual masking. For example, in the unidirectional mode, Sinkhorn Transformer sorts the buckets only according to the first token in each bucket. Bidirectional Clustered Attention could first approximate the full contextual attention scores with that of the cluster centroid each query belongs to, while separately computing on the top-$k$ keys. However, this technique is not viable in a unidirectional setting.
    \item \textit{Using a larger bucket size does not necessarily improve the performance.} We observe this phenomenon from the results of Sinkhorn Transformer and Routing Transformer. While the bucket size is increased, hence the context size for each query, we use fewer buckets/clusters. This would make it harder for k-means clustering and Sinkhorn sorting to group relevant items together. Hence, one has to carefully tune the bucket size to achieve ideal performance, as it balances between the size of attention context and the quality of sorting / bucket assignment. Surprisingly, we find using multiple rounds of hashing in Reformer does not enhance the performance either.
\end{itemize}

\subsection{Computational Cost}
\subsubsection{Settings}
\label{sec:exp_computation_cost}
To evaluate the computational efficiency of our proposed method, we compare the inference speed of our method with the vanilla Transformer and the aforementioned efficient attention baselines. Following~\cite{kitaev2019reformer,wang2020linformer}, we use synthetic inputs with varying lengths from 128 to 64K, and perform a full forward pass. The batch size is scaled inversely with the sequence length, to keep the total number of items (tokens) fixed. We report the average running time on 100 batches. For each baseline model, we only consider the less time-consuming variant. For example, we only report the LSH-1 variant for Reformer, as the LSH-4 version is far more computationally intensive. Since the asymptotic complexity of our proposed method is $O(LBW)$, the inference speed of all the three versions of LISA only depends on the total number of codewords used to encode sequences (i.e., $BW$). We evaluate two settings of LISA that use a total of 128 and 256 codewords (denoted as \textbf{LISA-128} and \textbf{LISA-256}), corresponding to the settings we used for LISA-Soft and LISA-Mini in Section~\ref{sec:rec_perf}. We only measure the cost of self-attention, since other components are the same for all compared models. We consider latent dimension sizes of 128 and 1024. All the experiments are conducted on a single Tesla V100 GPU with 16GB memory. The results are shown in Figure~\ref{fig:inference_speed}.

\subsubsection{Findings}
\begin{itemize}[leftmargin=*]
    \item \textit{Our method consistently and dramatically outperforms Transformer and all efficient attention baselines in inference speed.} Only on sequences of length 128 and using an embedding size of 128, LISA is slightly slower than Transformer. When $D=128$, LISA-128 is 3.1x faster than Reformer on 64K sequences. Benefiting from using inner product tables, our method is even way faster than others when $D=1024$, achieving a speed boost of 57x compared to Transformer on 16K sequences. All other methods take considerably longer time as the cost of computing the inner product dominates in this scenario. Linformer has an almost identical speed to LISA-128 when $D=128$. However, its recommendation performance is notably worse than ours. From Figure~\ref{fig:inference_speed}, we also verify the linear complexity of LISA, as the inference time remains constant when the total number of items in a batch is constant.
    \item \textit{Sinkhorn Transformer and Routing Transformer still suffer from enormous computational cost with growing sequence length.} Especially when $D=128$, the inference time increases by 5x for Sinkhorn and 27x for Routing moving from sequences of 128 to 64K. Both the two methods require $O(LB)$ time to compute query/key dot product within each bucket, where $B$ is the bucket size. Sinkhorn Transformer takes $O((L/B)^2)$ time to sort buckets, while Routing Transformer spends $O(L^2/B)$ time to perform cluster assignments. With the bucket size fixed, the cost of sorting/clustering becomes dominant. Increasing the bucket size, on the other hand, we would have to pay an extra price in computing attention scores within each bucket.
    \item \textit{Though the extra overhead dominates when sequences are short, Reformer tends to be almost linear when facing longer sequences.} We see that hashing items into buckets via LSH is exceptionally time-consuming. When $D=128$, Reformer is significantly slower than the vanilla Transformer on sequences shorter than 512, even slower than on 64K sequences due to larger batch size. Our method, on the contrary, does not suffer from this issue, being up to 6.5x faster than Reformer on sequences of 128. On longer sequences, Reformer scales almost linearly, since the term $\log L$ is quite small in its asymptotic complexity $O(L \log L)$.
    \item \textit{Clustered Attention fails to demonstrate its advantage of linear complexity even on sequences of 64K.} From Figure~\ref{fig:inference_speed}, we observe that the Clustered Attention is indeed linear (although bears the same extra overhead problem handling short sequences as Reformer). It seems that there underlies a substantial computational cost by computing full-contextual attention using the cluster centroids, and then improving the approximation for each query on the top-$k$ keys. Clustered Attention is still slower than Reformer on 64K sequences.
\end{itemize}

\subsection{Memory Consumption}
\subsubsection{Settings}
We also evaluate the memory efficiency of different methods by measuring the peak GPU memory usage. The settings of the compared methods are the same as the previous section's. The latent dimensionality $D$ is set to 128. For a given sequence length, we choose the batch size to be the maximum that all compared models can fit in memory. We report results on sequences up to 16K long, as the vanilla Transformer could not fit longer sequences even with a batch size of 1. The compression ratios with respect to Transformer are shown in Table~\ref{tab:memory_consumption}.

\begin{table}
    \caption{Memory efficiency of different methods. The numbers in the table are the ratios between the peak memory usage of the Transformer and that of the compared efficient attention method. Bold font denotes the most memory efficient one.}
    \label{tab:memory_consumption}
    \resizebox{0.98\columnwidth}{!}{
    \begin{tabular}{lcccccc}
        \toprule
        & \multicolumn{6}{c}{sequence length} \\
        \cmidrule(lr){2-7}
        & 512 & 1024 & 2048 & 4096 & 8192 & 16384 \\
        \midrule
        Linformer & 2.46x & 4.48x & 8.49x & 16.51x & 32.65x & 65.53x \\
        Reformer & 0.66x & 1.16x & 2.15x & 4.16x & 8.31x & 11.26x \\
        Sinkhorn & 0.97x & 1.70x & 3.15x & 6.09x & 12.14x & 25.74x \\
        Routing & 1.27x & 2.22x & 4.08x & 7.73x & 14.87x & 29.45x \\
        Clustered & 2.32x & 4.17x & 7.85x & 15.26x & 30.63x & 64.91x \\
        LISA-128 & \textbf{2.94x} & \textbf{5.14x} & \textbf{9.55x} & \textbf{18.45x} & \textbf{36.86x} & \textbf{78.26x} \\
        LISA-256 & 1.50x & 2.62x & 4.87x & 9.40x & 18.78x & 39.93x\\
        \bottomrule
    \end{tabular}
    }
\end{table}

\subsubsection{Findings}
All the efficient attention baselines greatly reduce the memory consumption on longer sequences. Among which LISA-128 is the most efficient one, requiring only 1.3\% of the memory needed by Transformer in the best case. Although Reformer enjoys faster inference speed on long sequences, we see that it is more memory-hungry than other baselines. This again reflects the LSH bucketing overhead of Reformer.

\begin{figure}
	\centering
	\includegraphics[width=.98\linewidth]{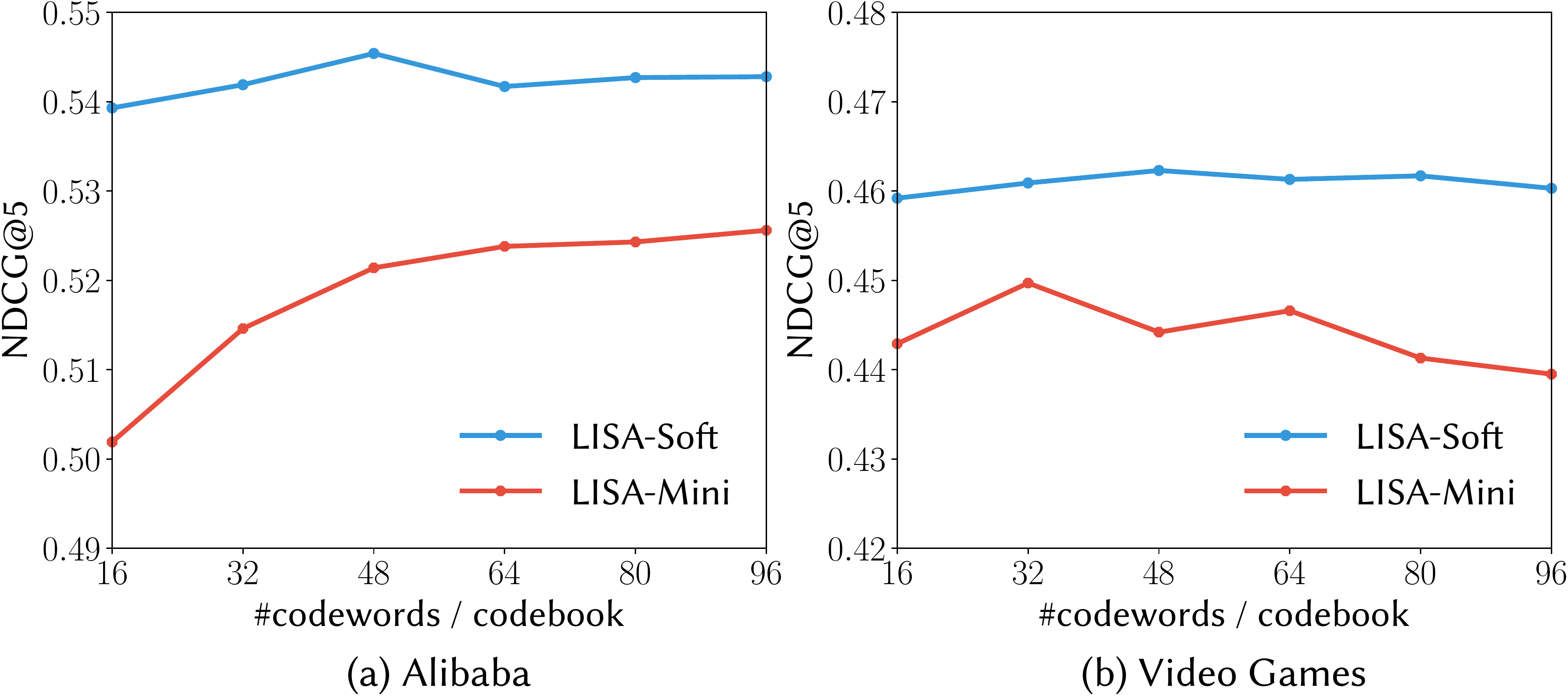}
	\caption{The impact of the number of codewords.}
	\label{fig:sensitivity_codewords}
\end{figure}

\subsection{Sensitivity w.r.t. Number of Codewords}
\subsubsection{Settings}
We investigate the impact of the number of codewords per codebook (i.e., $W$) used in LISA-Soft and LISA-Mini on recommendation performance. We keep the number of codebooks to be 8 and vary $W$ from 16 to 96. We leave the settings of the codebooks used to encode target items in LISA-Mini unchanged. We show the results on Alibaba and Video Games in Figure~\ref{fig:sensitivity_codewords}.

\subsubsection{Findings}
The performance of LISA-Mini on Alibaba consistently improves with the increasing number of codewords used. Due to the sparsity of Video Games, it is challenging to learn two large codebooks well simultaneously. Hence the performance drops a bit when using a large number of codewords on this dataset. On the other hand, the performance of LISA-soft is relatively stable w.r.t. $W$ on both datasets, indicating that we can attain desirable performance with only a small number of codewords, greatly boost the inference efficiency.

\begin{figure}
	\centering
	\includegraphics[width=.75\linewidth]{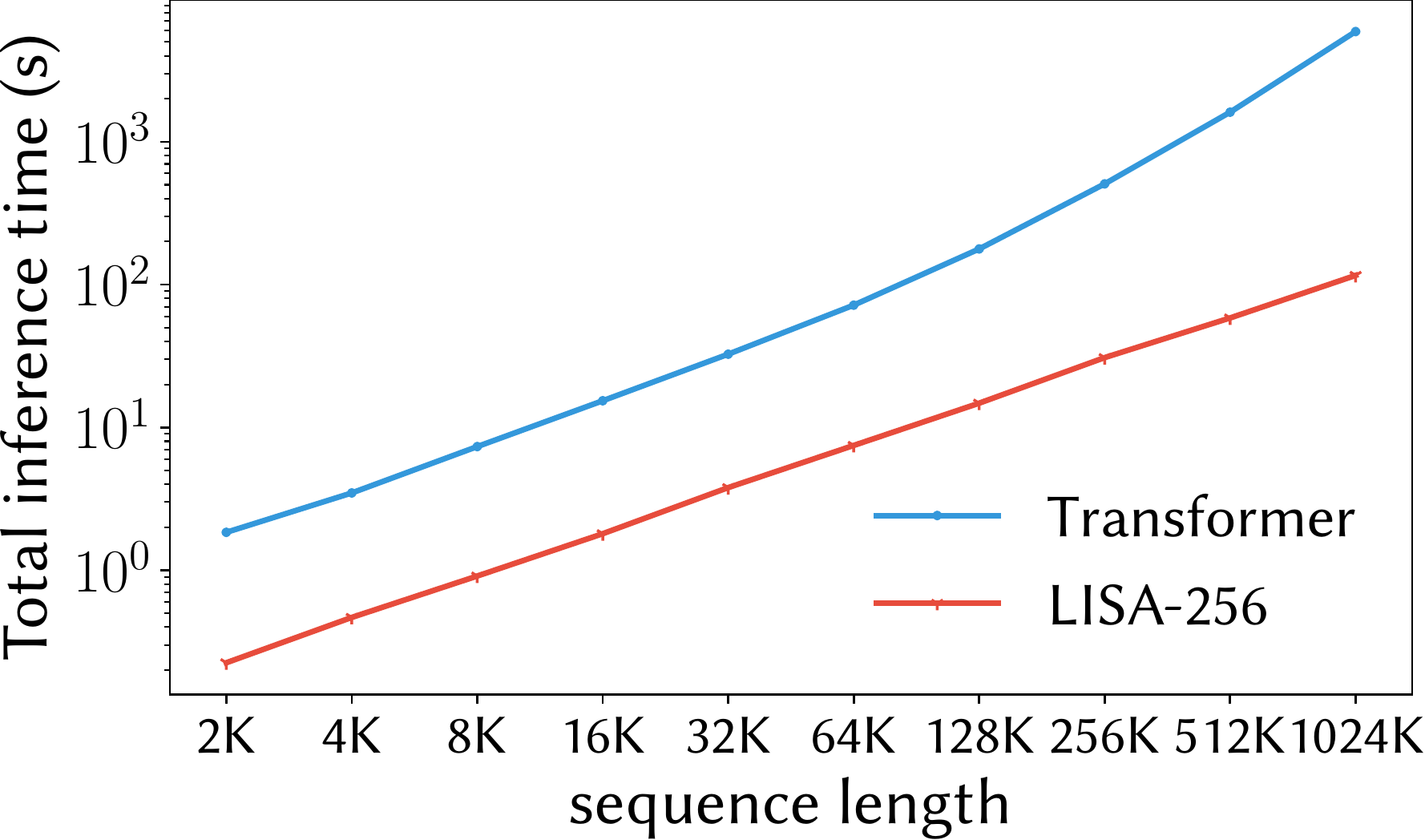}
	\caption{Inference speed of Transformer and LISA in an interactive setting.}
	\label{fig:stepwise_inference}
\end{figure}

\subsection{Improving Efficiency for Online Recommendation}
Here we consider a practical setting that users and the recommender interact in a dynamic manner. The recommender makes recommendations based on the user's historical behaviors. The user then interacts with the recommendations, and the response is appended to the user's history. This process is repeated as the recommender makes new recommendations using the updated user sequence.

A particular advantage of our method emerges in this setting. In our method, the computation of the attention scores only depends on the codeword histogram and the codebooks themselves. For each user, instead of having to store his entire history sequence at the cost of $O(L)$, we can just save the codeword histogram and the last item's codeword indices to represent the user's state. The codeword histogram and the indices can be dynamically updated, resulting in a constant storage cost of $O(BW)$. At each inference step, our method can utilize the stored histogram to compute a weighted average of codebooks in a constant time of $O(BWD)$, compared with the $O(LD)$ complexity for the vanilla self-attention.

We simulate this scenario with randomly generate data. The total time required to make stepwise inferences from scratch up to some length $L$ is measured. Since most efficient attention baselines face challenges when dealing with variable sequence length (recall that Sinkhorn Transformer and Linformer assume a fixed sequence length as their model parameters depend on this length), we only compare LISA-256 with the Transformer.

We show the results in Figure~\ref{fig:stepwise_inference}. We see that our method is considerably faster than Transformer in this setting, especially at a larger number of steps. Concretely, our method takes about 0.11ms to progress a step, no matter how long the sequence it. However, it would take Transformer 0.98ms to compute attention for a single query when the sequence is at 2K length, 1.50ms at 64K, and 11.01ms at 1024K, \textasciitilde 100x slower than our method.

\subsection{Migrating Codebooks from Vanilla Self-Attention}
\subsubsection{Settings}
We note that the codebooks serve as a plug and play module, which can be used to replace any embedding matrix. We can also train the model based on vanilla self-attention with codebooks. The pretrained codebooks are directly applied to LISA-Base and are frozen. We evaluate the performance of this model and report the results in Table~\ref{tab:distillation}. 

\subsubsection{Findings}
We find that directly use codebooks trained with regular dot-product attention does not cause performance degradation, but actually improves the performance of the LISA-Base model a little. This implies that our method indeed can approximate dot-product attention to some extent.

\begin{table}
    \caption{Performance of LISA-Base using codebooks pretrained with the vanilla Transformer.}
    \label{tab:distillation}
    \begin{tabular}{lcccc}
        \toprule
        & HR@5 & NDCG@5 & HR@10 & NDCG@10 \\
        \midrule
        Alibaba & 0.6697 & 0.5492 & 0.7711 & 0.5821 \\
        ML-1M & 0.7002 & 0.5456 & 0.7945 & 0.5763 \\
        Video Games & 0.6188 & 0.4800 & 0.7333 & 0.5172 \\
        ML-25M & 0.9287 & 0.7991 & 0.9725 & 0.8135 \\ 
        \bottomrule
    \end{tabular}
\end{table}
\section{Conclusions and Future Works}
In this paper, we propose LISA, an efficient attention mechanism for recommendation, built upon embedding quantization with codebooks. In LISA, codeword histograms for each codebook are computed over the input sequences. We then use the histograms and the inner product between codewords to compute the attention weights, in time linear in the sequence length. Our method performs on par with the vanilla Transformer in terms of recommendation performance, while being up to 57x faster. Future works can include extending LISA to other domains like language modeling. 

\section*{Acknowledgments}
The work was supported by grants from the National Natural Science Foundation of China (No. 62022077 and 61976198), and the Fundamental Research Funds for the Central Universities.

\bibliographystyle{ACM-Reference-Format}
\bibliography{reference}

\end{document}